\documentclass[11pt,a4paper]{article}
\usepackage[T1]{fontenc}
\usepackage{graphicx}
\usepackage{url}
\usepackage{color}
\usepackage{multirow}

\begin{document}
%

\title{Challenging Portability Paradigms: FPGA Acceleration Using SYCL and OpenCL\thanks{
\textit{Presented at HeteroPar 2024 (22nd International Workshop Algorithms, Models and Tools 
for Parallel Computing on Heterogeneous Platforms),}, Madrid, Aug. 27th 2024.
\vspace{0.5em}
}}
%
%


\author{
  Manuel de Castro\thanks{Dpt. of Computer Science, Univ. de Valladolid, Spain. manuel@infor.uva.es}, \and
  Roberto R. Osorio\thanks{CITIC, Comp. Arch. Group, Univ. da Coruña, Spain, roberto.osorio@udc.es},  \and
  Francisco J. Andújar\thanks{Dpt. of Computer Science, Univ. de Valladolid, Spain. fandujarm@infor.uva.es}, \and
  Rocío Carratalá-Sáez\thanks{Dpto. Ing. y Ciencia de los Computadores, Univ. Jaume I, Spain. rcarrata@uji.es}, \and
  Yuri Torres\thanks{Dpt. of Computer Science, Univ. de Valladolid, Spain. yuri.torres@infor.uva.es}, \and
  Diego R. Llanos\thanks{Dpt. of Computer Science, Univ. de Valladolid, Spain. diego.llanos@uva.es},
}

\date{}

\maketitle              
\begin{abstract}
As the interest in FPGA-based accelerators for HPC applications increases, new 
challenges also arise, especially concerning different programming and portability issues.
This paper aims to provide a snapshot of the current state of the FPGA tooling and its problems.
To do so, we evaluate the performance portability of two frameworks for developing 
FPGA solutions for HPC (SYCL and OpenCL) when using them to port a highly-parallel 
application to FPGAs, using both ND-range and single-task type of kernels. 

The developer's general recommendation when using FPGAs is to develop single-task kernels 
for them, as they are commonly regarded as more suited for such hardware. However,
we discovered that, when using high-level approaches such as OpenCL and SYCL to program
a highly-parallel application with no FPGA-tailored optimizations, 
ND-range kernels significantly outperform single-task codes. 
Specifically, while SYCL struggles to produce 
efficient FPGA implementations of applications described as single-task codes, 
its performance excels with ND-range kernels, a result that was unexpectedly favorable.



\vspace{0.5em}

\end{abstract}
%

\section{Introduction}
\label{sec:introduction}

In recent years, Field Programmable Gate Arrays (FPGAs) have gained popularity in 
High-Performance Computing (HPC) environments, as the slowdown in the increase of 
CPU performance incentivized research into hardware accelerators. In addition, with 
the end of Dennard scaling and Moore's law, power consumption has emerged as a major 
constraint of current-day high-performance systems. FPGAs present a special interest 
in this context, as they have proven to possess high power efficiency while being 
able to accelerate computationally costly tasks~\cite{deCastro2024a}.
Among others, FPGA accelerators have demonstrated their usefulness in many fields, such as Deep 
Learning, Finance, Signal and Multimedia Processing, and Fluid Dynamics.

FPGAs differ considerably from traditional load-store processor architectures (CPUs 
and GPUs) and these differences are responsible for making FPGAs competitive 
accelerators in certain fields~\cite{fpgas_dnns}. On the other hand, they also pose 
additional challenges and limitations for developing FPGA-targeting applications. 
Namely, efficient FPGA programming requires a deep knowledge of hardware concepts 
with which software developers are usually unfamiliar.
Hardware Description Languages 
(HDLs), such as VHDL and Verilog, have historically been 
used by electronic engineers to design and deploy custom hardware architectures on 
FPGAs. These languages differ significantly from the programming languages widespread 
among software developers, even in HPC contexts. 
High-Level Synthesis (HLS) frameworks, such as Vivado HLS~\cite{o2014xilinx},
have been developed to ease these programming complexities by allowing programmers 
to use dialects of high-level programming languages to target FPGAs. 
In general, even when using any of these HLS frameworks, efficient programming 
of FPGAs requires considerable additional development effort, which translates into
longer development cycles when compared to programming other accelerators, such as 
multicore CPUs with OpenMP, or GPUs with CUDA.

To ease the development process of HPC applications for different accelerators, 
two frameworks were proposed with portability in mind: OpenCL and SYCL. OpenCL~\cite{khronosopencl} 
is a programming framework for heterogeneous environments that can be used as an 
HLS development tool for both Intel and Xilinx FPGAs. Before integrating the FPGA support, 
OpenCL had already been widely adopted in HPC environments to program heterogeneous 
systems, and thus has a special interest for the HPC community as an HLS language. 
It is compatible with any C program, and seeks to provide code portability among 
many different computing units. OpenCL is mainly 
used in HPC environments to program data-parallel, Single Instruction Multiple Data 
(SIMD) accelerators, such as multicore CPUs and GPUs.
Although the SIMD-optimized codes could be compiled and run on an FPGA without any 
major change, additional development is most likely required to efficiently exploit 
the FPGA's resources and achieve a performance-competent implementation.

Similarly, SYCL~\cite{SYCL_2020} is a C++ library and abstraction layer 
for heterogeneous computing that can target FPGA accelerators (using Intel's DPC++ 
SYCL implementation). SYCL is built on top of OpenCL to provide a generic and portable 
programming framework for heterogeneous systems, leveraging C++ specific features, 
such as templates and generic lambda functions, to provide a higher level of abstraction. 
Regarding its usage for programming FPGAs, SYCL presents issues similar to those 
of OpenCL: SYCL is an abstraction layer that mainly focuses on data-parallel approaches, 
and efficiently programming FPGAs usually requires additional code transformations 
and efforts.

In this work, we evaluate and compare SYCL and OpenCL, two commonly-used HPC heterogeneous 
programming tools for developing applications on Intel FPGAs. 
To that end, we have chosen to use these frameworks to port UVaFTLE~\cite{rocio_uvaftle}, a highly-parallel 
fluid dynamics calculus to extract the
Finite Time Lyapunov (FTLE) exponent for fluid dynamic applications. 
UVaFTLE is a set of modern, open-source multicore 
and multi-GPU implementations of the FTLE computation using OpenMP and CUDA. 
The FTLE computation is an embarrassingly parallel task, as it involves performing 
multiple computations over each of the flowmap's particles independently. 

The migration of FPGA codes to SYCL has not been explored in the literature. Instead, 
some works focus on the design of FPGA hardware accelerators from the ground up using 
SYCL, such as in~\cite{sycl_fpga}. To the best of our knowledge, no previous works 
explore the performance differences between ND-range and single-task FPGA kernels 
when using SYCL.
More works study the use of OpenCL to develop HPC solutions for Intel FPGAs. Some 
study the optimization of well-known HPC kernels to 
FPGAs~\cite{zohouri_opencl_evaluation,opencl_analysis_framework}, 
with some of them exploring the performance differences between ND-range and single-task 
kernel versions of the same algorithms~\cite{hassan_dwarfs,verma_dwarfs}. 
While some of these works provide performance comparisons against 
GPU implementations, none of them explore the performance comparison against other 
FPGA implementations using different frameworks. 

Our results show that the use of ND-range kernels to deploy our
application in a FPGA offers much better performance than the use of single-task 
kernels. This suggests that single-task kernel performance cannot be improved without
applying optimizations specifically tailored for FPGAs, thus requiring some knowledge
that is usually outside the expertise of HPC programmers. All
the codes and performance analysis carried out in this work are publicly available
at \texttt{https://github.com/uva-trasgo/uvaftle/tree/fpga}.

\section{The problem: FTLE}
\label{sec:ftle}

The field of fluid dynamics has been widely explored from the computational perspective 
because of its importance in a wide variety of engineering applications. One of the 
topics that are of great interest in this field is the determination of the fluid 
particle trajectories in phase space, also known as {\it Lagrangian particle dynamics}, and the
calculation of the corresponding \textit{Lagrangian Coherent Structures (LCS)}~\cite{Haller2015}. 
LCS are recognized as influential surfaces within a dynamic 
system, guiding the paths of nearby trajectories throughout a specific period of interest. 
They play a pivotal role in directing the movement and formation of various physical occurrences 
such as oceanic entities like floating debris or oil slicks, as well as atmospheric phenomena 
including volcanic ash clouds and spore dispersals. 
%
The main interest in computing the LCS is because they provide a better understanding 
of the flow phenomena, since they can be broadly interpreted as \textit{transport 
barriers} that influence the transport within the flow.

LCS are not directly observable but can be inferred through the calculation of FTLE (Finite Time
Lyapunov Exponents) fields.
The fluid particle trajectories are defined as solutions of 
$\dot{\vec x} = \vec v \left(\vec x,\, t \right)$,
where the right-hand side is the velocity field of the fluid, in the absence of molecular 
diffusion. Solving this system of equations allows the LCS to be calculated. From 
the computational point of view, the extraction of LCS is achieved by completing 
 two main steps: The flowmap computation and the FTLE extraction. In this paper, 
we take the second step as our target calculation: See~\cite{Brunton2009} for the 
mathematical details. 

In our previous 
work, we presented UVaFTLE~\cite{rocio_uvaftle}, an optimized HPC application for computing 
the FTLE, given the description of a fluid's flowmap, using multi-core CPUs (using 
OpenMP), GPUs (using CUDA), or a combination of both. 
The UVaFTLE application includes two kernels to compute the FTLE of a 2D or 3D flowmap, 
respectively. These kernels are executed by the accelerators to achieve high performance, 
and both are used in the present work. UVaFTLE also includes an additional 
computationally-costly kernel that preprocesses the provided input set before the 
FTLE computation step. Nevertheless, the acceleration of this latter kernel on FPGAs 
is beyond the scope of this work.


It is worth mentioning that we have also explored in~\cite{arxiv-UVaFTE-SYCL} the 
portability of the code by using SYCL to target heterogeneous GPU environments, whose 
basic SYCL kernels served as the starting point for this work.

\section{Development tools and algorithmic strategies}
\label{sec:frameworks}

We evaluate FPGA implementations of UVaFTLE in OpenCL (using Intel FPGA SDK for OpenCL) 
and SYCL (using Intel DPC++). Both frameworks are high-level abstractions for heterogeneous 
computing widely used in HPC environments. The main difference between them is that 
SYCL is C++-based and frequently uses C++ features, such as templates and generic lambda 
functions, to achieve a higher level of abstraction, whereas OpenCL is compatible 
with any C environment and provides lower-level control to the user, but it is consequently 
more verbose. Internally, SYCL can target multiple, different backends to compile 
a single generic source code, automatically optimizing the code for that backend 
in the process, thus maximizing the achievable 
performance in highly-heterogeneous systems. To optimize and transform the generic 
codes for the different backends, and to allow for single-source host-and-device 
codes, SYCL requires the use of a special compiler. OpenCL, on the other hand, can 
be used as a standalone C runtime library: It achieves genericity by compiling the 
codes at execution time for the specific accelerators present in the target machine. 
However, in the case of FPGA kernels, 
a special compiler is also needed to compile the device 
codes of the OpenCL implementations; and it is necessary to keep the host code and device code 
in separate files. Both frameworks internally leverage Intel's FPGA compiler
to synthesize the high-level designs into hardware architectures.


Both SYCL and OpenCL allow two different ways of writing computational kernels to 
be deployed to accelerators. One is known as \textit{ND-range} kernels; which is the most 
common way of writing SYCL and OpenCL kernels. The other is known as \textit{single-task} 
kernels. 

ND-range kernels use a data-parallel approach. Conceptually, the workload is divided 
into multiple \textit{threads}, thus each performs a fraction of the work. The kernel 
code is written at the single-thread level; that is, the code describes the operations that 
each of the ND-range threads should perform individually. All threads are assigned 
a global identifier by the runtime system, which can be used to describe divergences 
among different threads in the code. In SYCL and OpenCL terminology, the threads 
that execute the kernel are called \textit{work-items}, which are at the same time 
grouped into \textit{work-groups}. An \textit{ND-range} is an N-dimensional indexed
space, where N can be one, two, or three. The identifier of each work-item is based 
on its n-dimensional coordinates within the index space. The core idea behind this 
design philosophy is that work-items may be executed in parallel, with up to one 
work-group of work-items being executed at the same time. Nevertheless, that is just a 
conceptual model that is not guaranteed to describe the actual underlying kernel 
execution process: The runtime abstracts the low-level details of the real execution 
model.

Regarding single-task kernels, they are also known as \textit{single work-item} kernels in SYCL and 
OpenCL terminology. That is because they are written to be executed using just one 
work-item (thread). They are semantically equivalent to ND-range kernels that are 
executed with just one work group of one work-item. Thus, single-task kernels are 
written at a global-problem level, and require explicit management of any possible 
parallelism (thread or otherwise).


SIMD-like data-parallel accelerators, such as multicore CPUs and GPUs, benefit greatly 
from ND-range kernels, since these kernels efficiently map to their massively parallel 
architectures. ND-range kernels on these accelerators enable high-performance gains 
with relatively low development costs. Single-task kernels are rarely ever used together 
with SIMD-like accelerators, since their achievable performance is rather poor in 
comparison. In these contexts, single-task kernels are usually relegated to corner 
cases; namely when a single-threaded task of low computational complexity is to be 
executed in-device, instead of in the host side of the system.

FPGA accelerators, however, do not share the characteristics of SIMD-like accelerators. 
Instead, they attempt to accelerate computational tasks by building tailored solutions 
(hardware architectures) out of available resources (low-level electronic components), 
laid out and interconnected in an abstract, limited space (the FPGA fabric or ``area''). 
As FPGA vendors point out~\cite{intel_fpga_description}, FPGAs can exploit multiple 
types of parallelism (including SIMD parallelism and task parallelism); nevertheless, 
the highest performance gains are usually achieved through pipeline parallelism. 
FPGAs can instantiate deep pipelines, which, when fully occupied, achieve a high 
performance. FPGA pipelines can be significantly deeper and more specialized than 
those of fixed, general-purpose computing units (CPUs and GPUs). 
Performance can also be further optimized 
by combining other kinds of parallelism, e.g. by designing the pipeline in a SIMD 
fashion, where multiple data are processed at the same time on every pipeline stage, 
or by instantiating multiple pipelines at the same time to achieve task parallelism.

\enlargethispage{1mm}
With all this in mind, both the Intel FPGA SDK for OpenCL, and the DPC++ programming 
frameworks point out in their guides that kernels that perform favorably on FPGAs 
are often expressed as single-task kernels~\cite{intel_fpga_description}. In both environments, 
the general recommendation is to write the computing kernels as single-task kernels 
whenever possible, unless the kernel maps naturally and exceptionally well to an 
ND-range structure~\cite{prefer_single-task}. However, as we will see in the following 
sections, we have found
that ND-range kernels perform consistently better than single-task approaches for
our application.

\section{Naïve implementation approach}
\label{sec:naive}

In the original UVaFTLE paper~\cite{rocio_uvaftle}, we provided open-source 
implementations of the application targeting multicore CPUs (using OpenMP) and 
GPUs (using CUDA). As stated previously, the FTLE computation is an embarrassingly 
parallel problem. 
Therefore, these original implementations were able to achieve high performance, 
as shown in~\cite{rocio_uvaftle}. Moreover, 
the OpenMP and CUDA implementations can be used to make naïve ports of the application 
to FPGA execution frameworks.

We used the CUDA version shown in~\cite{rocio_uvaftle} as a baseline for the 
naïve ND-range FPGA kernels, since 
CUDA programming follows the ND-range model of SYCL and OpenCL. The OpenMP version 
was used as a baseline for the naïve single-task kernels, as it more closely resembles 
a sequential version of the application and, consequently, can be more easily used 
to develop a single-task-type kernel. 
The naïve kernels developed represent low development-effort ports of the application 
to FPGAs; although, as noted in many works such as~\cite{hassan_dwarfs}, the performance 
achieved by these ports may not be competitive.

Regarding the host side of the code, naïve SYCL implementations were derived from 
the GPU SYCL implementations developed for~\cite{arxiv-UVaFTE-SYCL}. Being SYCL a 
high-level programming model designed with heterogeneous portability in mind, the 
changes required to port the GPU implementations to FPGAs are minimal. It is worth 
noting that the SYCL implementation used in that work, AdaptativeCpp (formerly OpenSYCL 
/ HipSYCL), cannot target Intel FPGAs; thus, DPC++ is used as the SYCL implementation 
of choice. This entails additional modifications of the code (namely namespaces and 
headers), yet the code modifications are still minimal.

The host code for the OpenCL implementations can be easily derived from the CUDA 
host code by changing the CUDA API calls to the equivalent OpenCL API calls, since 
the two programming models are very similar.

\begin{table}[t]
\centering
\scriptsize
\begin{tabular}{|c||c|c|c||c|c|c|}
\hline
\multirow{2}{*}{\textbf{Implementation}} & \multicolumn{3}{c||}{\textbf{~2D version (\# points)~}} & \multicolumn{3}{c|}{\textbf{3D version (\# points)}} \\
\cline{2-7}
& \textbf{200K} & \textbf{400K} & \textbf{600K} & \textbf{200K} & \textbf{400K} & \textbf{600K} \\
\hline
S-NR naïve & 11.1 & 22.3 & 33.7        & 371.7 & 803.1 & 1\,364.9 \\ \hline      
O-NR naïve & 10.7 & 21.5 & 32.5        & 359.4 & 777.6 & 1275.7 \\ \hline
S-ST naïve & 6\,635.5 & 13\,316.5 & 6\,281.4 & 26\,892.7 & 54\,207.4 & 34\,863.8 \\ \hline
O-ST naïve & 2\,085.7 & 4\,116.1 & 20\,034.4 & 9\,194.2 & 18\,455.6 & 92\,703.2 \\ \hline
CPU 1 thread & 30.1 & 51.1 & 75.5   & 71.5 & 172.6 & 240.1 \\ \hline
CPU 4 threads & 20.7 & 32.5 & 39.0  & 41.3 & 64.0 & 90.1 \\ \hline
CPU 8 threads & 17.2 & 23.1 & 31.6   & 33.0 & 51.3 & 70.6 \\ \hline    
\end{tabular}

\vspace{\baselineskip}
\caption{Execution times, in milliseconds, for the naïve implementations of the 2D and 3D
FTLE kernels for FPGA, and a reference CPU implementation parallelized using OpenMP. 
Legend is as follows: S- and O- are for SYCL and OpenCL, respectively; NR and ST 
are for ND-range and single-task, respectively.}
\label{tab:exp_naive_2D_3D}
\end{table}
\enlargethispage{2mm}
\section{Analysis of the results obtained}
\label{sec:results}

Based on those results described in the last sections, we can now make a series of observations.

\begin{enumerate}

\item Across all experiments, ND-range kernels consistently 
surpassed the performance of single-task kernels, albeit by a marginal difference in scenarios 
involving optimized kernels. Notably, the naïve implementations of single-task versions, 
especially those using SYCL, exhibited exceptionally poor performance.


\item The naïve ND-range implementations deliver robust performance right out of the gate. While 
not reaching the heights of optimized gradient kernels, their performance closely aligns with the 
total execution times observed in the optimized hardware-software approach for FTLE 
computations, which includes both CPU and FPGA components. Consequently, we conclude that 
ND-range kernels exhibit remarkable portability across both SYCL and OpenCL platforms, 
particularly for embarrassingly parallel tasks.


\item The optimized FPGA kernels approach the theoretical peak performance for gradient 
computation, processing nearly one point per FPGA clock cycle. This performance suggests that 
SYCL and OpenCL compilers are capable of inferring architectures that are nearly optimal, 
given a well-constructed kernel.


\item In comparisons between SYCL and OpenCL, OpenCL consistently outperformed SYCL across 
all tests, highlighting a performance trade-off associated with SYCL's higher abstraction levels. 
This gap is particularly pronounced in single-task kernels, where SYCL lags significantly behind 
OpenCL. Despite Intel discontinuing the FPGA SDK for OpenCL in August 2023 and recommending 
SYCL for FPGA development, the superior performance of OpenCL may still make it a preferred 
choice for developers, especially for single-task kernels --- Intel's recommended kernel type. 
While OpenCL also edges out SYCL in ND-range kernel performance, the difference is marginal.

\end{enumerate}


\section{Conclusions}
\label{sec:conclusions}



In conclusion, our findings reveal that naïve ND-range codes exhibit remarkable portability 
for highly parallel applications, in stark contrast to the significantly lower portability of 
single-task codes. Moreover, we observed that SYCL faces challenges in generating efficient 
hardware architectures for applications characterized as single-task codes, resulting in 
performance that is up to three orders of magnitude inferior compared to other implementations. 
Intriguingly, SYCL's performance dramatically improves when employing the ND-range approach, 
a result that surpasses expectations.

Future work includes to explore whether the use of more FPGA-specific optimizations in single-task
kernels may alleviate the lack of performance of this approach. We also plan to 
perform experimental evaluations of equivalent Vivado HLS codes 
on Xilinx FPGAs, identifying the reasons behind the much lower working frequencies 
than the ones reported by Vivado for an equivalent implementation, and studying the 
optimal approach to accelerate the determination of the list of neighbors for each point.

\subsubsection{Acknowledgments}
This work was supported in part by: the Spanish Ministerio de Ciencia e Innovación 
and by the European Regional Development Fund (ERDF) program, 
under Grant PID2022-142292NB-I00 (NATASHA Project); 
and in part by the Junta de Castilla 
y León - FEDER Grants, under Grant VA226P20 (PROPHET-2 Project), Junta de Castilla 
y León, Spain. 
This work was also supported in part by grant TED2021–130367B–I00, 
funded by MCIN/AEI/10.13039/ 501100011033, by “European Union NextGenerationEU/PRTR”; 
by grant EDC431C 2021/30 (Xunta de Galicia, Consolidation Program of Competitive
Reference Groups);
and by grant PID2022-136435NB-I00, funded by MCIN/AEI/ 10.13039/501100011033 and 
by "ERDF A way of making Europe", EU. 
Manuel de Castro has been supported by Spanish Ministerio de
Ciencia, Innovación y Universidades, through ``Ayudas para la Formación
de Profesorado Universitario FPU 2022''.
%
The authors have no competing interests that
might be perceived to influence the results and/or discussion reported in this paper.

\bibliographystyle{splncs04}

\begin{thebibliography}{10}
\providecommand{\url}[1]{\texttt{#1}}
\providecommand{\urlprefix}{URL }
\providecommand{\doi}[1]{https://doi.org/#1}

\bibitem{Brunton2009}
Brunton, S., Rowley, C.: Modeling the unsteady aerodynamic forces on
  small-scale wings. In: 47th AIAA Aerospace Sciences Meeting. p.~1127 (2009).
  \doi{10.2514/6.2009-1127}

\bibitem{arxiv-UVaFTE-SYCL}
Carratalá-Sáez, R., et~al.: {Open SYCL} on heterogeneous {GPU} systems: A
  case of study. ArXiv preprint pp. 0--17 (2023).
  \doi{10.48550/arXiv.2310.06947}

\bibitem{rocio_uvaftle}
Carratalá-Sáez, R., et~al.: {UVaFTLE: Lagrangian finite time Lyapunov
  exponent extraction for fluid dynamic application}. {The Journal of
  Supercomputing}  \textbf{79},  9635--9665 (2023).
  \doi{10.1007/s11227-022-05017-x}

\bibitem{deCastro2024a}
de~Castro, M., Vilariño, D.L., Torres, Y., Llanos, D.R.: The role of
  field-programmable gate arrays in the acceleration of modern high-performance
  computing workloads. Computer  \textbf{57}(7),  66--76 (2024).
  \doi{10.1109/MC.2024.3378380}

\bibitem{Haller2015}
Haller, G.: Lagrangian coherent structures. Annual Review of Fluid Mechanics
  \textbf{47},  137--162 (2015). \doi{10.1063/1.3690153}

\bibitem{hassan_dwarfs}
Hassan, M.W., et~al.: Exploring {FPGA}-specific optimizations for irregular
  {OpenCL} applications. In: ReConFig 2018. pp.~1--8.
  \doi{10.1109/RECONFIG.2018.8641699}

\bibitem{intel_fpga_description}
Intel: {Compare Benefits of CPUs, GPUs, and FPGAs for Different oneAPI Compute
  Workloads} (2022), \url{https://tinyurl.com/2zevt9p2}, (accessed March 12,
  2024)

\bibitem{prefer_single-task}
Intel: {Intel FPGA SDK for OpenCL™ Pro Edition: Best Practices Guide: 1.3
  Single Work-Item Kernel versus NDRange Kernel} (2022),
  \url{https://tinyurl.com/23amr92d}, (accessed March 12, 2024)

\bibitem{sycl_fpga}
Kamalakkannan, K., et~al.: {FPGA} acceleration of structured-mesh-based
  explicit and implicit numerical solvers using {SYCL}. In: IWOCL'22. ACM, New
  York, NY, USA (2022), \url{https://doi.org/10.1145/3529538.3530007}

\bibitem{SYCL_2020}
{Khronos {OpenCL} working group}: {SYCL 2020 Specification (revision 7)}
  (2020), \url{https://tinyurl.com/yc7kwb9u}, (accessed June 20, 2023)

\bibitem{khronosopencl}
{Khronos OpenCL Working Group}, et~al.: {The OpenCL Specification, version 1.0.
  29, 8 December 2008}, \url{https://www.khronos.org/opencl/}, accessed July
  2022

\bibitem{fpgas_dnns}
Nurvitadhi, E., et~al.: Can {FPGAs} beat {GPUs} in accelerating next-generation
  deep neural networks? pp. 5--14 (02 2017). \doi{10.1145/3020078.3021740}

\bibitem{o2014xilinx}
O'Loughlin, D., et~al.: Xilinx {Vivado} high level synthesis: Case studies. In:
  25th ISSC2014/CIICT2014. IET (2014). \doi{10.1049/cp.2014.0713}

\bibitem{verma_dwarfs}
Verma, A., et~al.: Accelerating workloads on {FPGAs} via {OpenCL}: A case study
  with {OpenDwarfs}. Tech. rep., Dept. of Computer Science, Virginia Tech
  (2016)

\bibitem{opencl_analysis_framework}
Wang, Z., et~al.: A performance analysis framework for optimizing {OpenCL}
  applications on {FPGAs}. In: HPCA 2016. pp. 114--125.
  \doi{10.1109/HPCA.2016.7446058}

\bibitem{zohouri_opencl_evaluation}
Zohouri, H.R., et~al.: Evaluating and optimizing {OpenCL} kernels for high
  performance computing with {FPGAs}. In: SC '16. pp. 409--420.
  \doi{10.1109/SC.2016.34}

\end{thebibliography}

\end{document}